\documentclass[12pt,preprint]{aastex}

\def\mearth{M_\oplus}

\slugcomment{Submitted to The Astrophysical Journal}

\shorttitle{Minimum masses of heavy elements in the envelopes of Jupiter and Saturn}
\shortauthors{O. Mousis et al.}

\begin{document}


\title{Determination of the minimum masses of heavy elements in the envelopes of Jupiter and Saturn\\
}


\author{
Olivier~Mousis\altaffilmark{1,2,3},
Ulysse~Marboeuf\altaffilmark{2,3},
Jonathan~I.~Lunine\altaffilmark{1},
Yann~Alibert\altaffilmark{2},
Leigh~N.~Fletcher\altaffilmark{4},
Glenn~S.~Orton\altaffilmark{4},
Fran\c coise~Pauzat\altaffilmark{5}
\& Yves~Ellinger\altaffilmark{5}}

\altaffiltext{1}{Lunar and Planetary Laboratory, University of Arizona, Tucson, AZ, USA.}

\altaffiltext{2}{Institut UTINAM, CNRS-UMR 6213, Observatoire de Besan\c con, BP 1615, 25010 Besan\c{c}on Cedex, France.}

\email{mousis@lpl.arizona.edu}

\altaffiltext{3}{The HOLMES collaboration.}

\altaffiltext{4}{Jet Propulsion Laboratory, California Institute of Technology, 4800 Oak Grove Drive, Pasadena, CA, 91109, USA.}

\altaffiltext{5}{Laboratoire de Chimie Th{\' e}orique (LCT/LETMEX), CNRS-UMR 7616, Universit{\' e} Pierre et Marie Curie, 4, Place Jussieu, 75252, Paris cedex 05 - France.}

\begin{abstract}

We calculate the minimum mass of heavy elements required in the envelopes of Jupiter and Saturn to match the observed oversolar abundances of volatiles. Because the clathration efficiency remains unknown in the solar nebula, we have considered a set of sequences of ice formation in which the fraction of water available for clathration is varied between 0 and 100\%. In all the cases considered, we assume that the water abundance remains homogeneous whatever the heliocentric distance in the nebula and directly derives from a gas phase of solar composition. Planetesimals then form in the feeding zones of Jupiter and Saturn from the agglomeration of clathrates and pure condensates in proportions fixed by the clathration efficiency. A fraction of Kr and Xe may have been sequestrated by the H$_3^+$ ion in the form of stable XeH$_3^+$ and KrH$_3^+$ complexes in the solar nebula gas phase, thus implying the formation of at least partly Xe- and Kr-impoverished planetesimals in the feeding zones of Jupiter and Saturn. These planetesimals were subsequently accreted and vaporized into the hydrogen envelopes of Jupiter and Saturn, thus engendering volatiles enrichments in their atmospheres, with respect to hydrogen. Taking into account both refractory and volatile components, and assuming plausible molecular mixing ratios in the gas phase of the outer solar nebula, we show that it is possible to match the observed enrichments in Jupiter and Saturn, whatever the clathration efficiency. Our calculations predict that the O/H enrichment decreases from $\sim$ 6.7 to 5.6 times (O/H)$_{\odot}$ in the envelope of Jupiter and from 18.1 to 15.4 times (O/H)$_{\odot}$ in the envelope of Saturn with the growing clathration efficiency in the solar nebula. As a result, the minimum mass of ices needed to be injected in the envelope of Jupiter decreases from $\sim$ 14.9 to 12.6 $\mearth$, including a mass of water diminishing from 9.6 to 7.7  $\mearth$. In the same conditions, the minimum mass of ices needed in the envelope of Saturn decreases from $\sim$ 12.2 to 10.6 $\mearth$, including a mass of water diminishing from 7.7 to 6.4 $\mearth$. The accretion of planetesimals with ices to rocks ratios $\sim$ 1 in the envelope of Jupiter, namely a value derived from the bulk densities of Ganymede and Callisto, remains compatible with the mass of heavy elements predicted by interior models. On the other hand, the accretion of planetesimals with similar ice--to--rock in the envelope of Saturn implies a mass of heavy elements greater than the one predicted by interior models, unless a substantial fraction of the accreted rock sedimented onto the core of the planet during its evolution.

\end{abstract}





\section{Introduction}

Measurements by the mass spectrometer aboard the Galileo probe have shown that the abundances of C, N, S, Ar, Kr and Xe are all enriched by similar amounts with respect to their solar abundances in the atmosphere of Jupiter (Mahaffy et al. 2000; Wong et al. 2004). Moreover, recent Cassini CIRS observations have also confirmed what was previously inferred from ground-based measurements,  that C is substantially enriched in the atmosphere of Saturn (Flasar et al. 2005; Fletcher et al. 2008).

In order to interpret these enrichments, it has been proposed by Gautier et al. (2001a,b), Alibert et al. (2005a,b), Mousis et al. (2006) and Hersant et al. (2004,2008) that the main volatile compounds initially existing in the solar nebula gas phase were essentially trapped by crystalline water ice in the form of clathrates or hydrates in the feeding zones of Jupiter and Saturn. These ices then agglomerated and formed planetesimals that were ultimately accreted by the forming Jupiter and Saturn. This is the fraction of these icy planetesimals that vaporized when entering the envelopes of the two growing planets which engendered the observed volatiles enrichments.

All the authors cited above postulate a full efficiency for the clathration process in the solar nebula, implying that guest molecules had the time to diffuse through porous water-ice planetesimals in the solar nebula. This remains plausible only if collisions between planetesimals have exposed essentially all the ice to the gas over time scales shorter or equal to planetesimals lifetimes in the nebula (Lunine \& Stevenson 1985). However, the efficiency of collisions between planetesimals to expose all the ``fresh'' ice over such a time scale still remains questionable and we have no evidence that clathration was important in the primordial nebula (Owen et al. 1999,2000). Moreover, in all the afore-mentioned works, the abundance of available water ice is considered as a free parameter to select the amount of volatile species that are fully enclathrated and that contribute to the observed enrichments. As a result, this leads to abundances of available water in the feeding zones of Jupiter and Saturn which depart significantly from that predicted from a solar composition gas ,and the validity of such choices still needs to be investigated.

Here, we show that it is possible to explain the volatiles enrichments in Jupiter and Saturn by postulating an incomplete clathration process in the solar nebula, and also by using an abundance of available water derived from a homogeneous gas phase of solar composition. In the extreme case of no clathration in the outer nebula, it is still possible to match the enrichments observed in the two planets. Volatile species that are not enclathrated essentially form pure condensates at lower temperatures in the feeding zones of Jupiter and Saturn and icy planetesimals ultimately accreted by the two forming planets then agglomerate from clathrates and pure condensates in proportions fixed by the clathration efficiency. We finally show that the minimum amounts of ices needed in the envelopes of Jupiter and Saturn to match their observed enrichments are compatible with the amounts of heavy elements predicted by interior models.

Sec. \ref{deli} is devoted to the description of the formation mechanisms of Jupiter and Saturn in the framework of the core-accretion model. We also briefly review the distribution of heavy elements in the two planets which is predicted by interior models. In Sec. \ref{ices}, we describe the formation sequence of icy planetesimals in the feedings zones of Jupiter and Saturn as a function of the clathration efficiency.
In Sec. \ref{mass}, we determine the minimum masses of heavy elements required to match the oversolar abundances of volatiles observed in Jupiter and Saturn. Sec. \ref{discus} is devoted to discussion.

\section{Delivery of volatiles to the envelopes of Jupiter and Saturn}
\label{deli}

There are  two main planet formation models considered in the recent literature. The first one, the disk instability model, postulates that planets form by gravitational instability in a protoplanetary disk. In the second model, the core-accretion model, planets form in two succesive phases. A solid core is first formed by collisional accretion of  planetesimals, in a process similar to the formation of terrestrial planets. In a second phase, when the mass of this core becomes large enough, rapid gas accretion is triggered, leading to the
formation of a gas giant planet. Recent improvements in core-accretion models (see e.g. Alibert et al. 2005c; Hubickyj et al. 2005) allow the formation of gas giant on time scales compatibles with disk observed lifetimes. Moreover, by including migration and disk evolution, some of these models allow forming planets whose internal structures are comparable with the ones of Jupiter and Saturn (Alibert et al. 2005b). Finally, in a population synthesis approach, the same model deliver planet statistical properties consitent with observed ones (at least for single planets in quasi-circular orbits around G stars; see e.g. Mordasini et al. 2008).

Our present work is based on the afore-mentioned extended core accretion model, taking into account migration and disk evolution. In this model, the two planets start their formation at larger distance from the Sun, migrate and simultaneously accrete gas and solids. When the disk has disappeared, the accretion stops, and the two planet do not migrate anymore. In this model, the structure of the protoplanetary disk is derived assuming that viscous heating is the predominant heating source. Under this hypothesis, disk models show that the outer parts of the disk are protected from solar irradiation by
shadowing effect of the inner disk parts, and temperature in the planet forming region (between 5 and 15 AU) decreases down to very low values ($\sim$ 20 K). However, note that irradiation onto the central disk parts could modify the disk structure so much that shadowing effect would be prevented in the outer parts. In this case, the temperature in the planet--forming region would be higher.

In this paper, we compare the amount of heavy elements needed to explain the enrichments in volatiles species in Jupiter and Saturn with theoretical determination of the planets metal content. In this context, we use the internal structure models of Saumon \& Guillot (2004) (hereafter SG04) who have derived estimates for the core mass and mean metallicity of Jupiter and Saturn, the main uncertainty in these determinations being the equation of state of hydrogen and helium under high pressure. Taking into account these different uncertainties, the total mass of heavy elements present in Jupiter ($M_{\rm core}$ + $M_{\rm Z, enve}$) can be as high as $\sim$ 42 $\mearth$ whereas the mass of the core can range between 0 and 13 $\mearth$. In the case of Saturn, the mass of heavy elements can increase up to $\sim$ 35 $\mearth$ with the envelope mass varying between $\sim$ 0 and 10 $\mearth$ and the core mass ranging between $\sim$ 8 and 25 $\mearth$. Note that these estimates will be improved, in the case of Jupiter, by the future JUNO mission.

\section{Formation sequence of icy planetesimals in the giant planets feeding zones}
\label{ices}

We describe here the formation sequence of the different ices produced in the feeding zones of proto-Jupiter and proto-Saturn. Once formed, these ices will add to the composition of the planetesimals that will be accreted by the two giant planets during their growth.

In order to define the initial gas phase composition of the solar nebula, we assume that the abundances of all elements, including oxygen, are protosolar (Lodders 2003) and consider both refractory and volatile components. Refractory components include rocks and organics. According to Lodders (2003), rocks contain $\sim$ 23\% of the total oxygen in the nebula. The fractional abundance of organic carbon is assumed to be 55\% of total carbon (Pollack et al.\ 1994), and the ratios of C/O/N included in organics is assumed to be 1/0.5/0.12 (Jessberger et al.\ 1988). We then assume that the remaining O, C, and N exist only under the form of H$_2$O, CO, CO$_2$, CH$_3$OH, CH$_4$, N$_2$, and NH$_3$.  Hence, once the gas phase abundances of elements are defined, the abundances of CO, CO$_2$,  CH$_3$OH, CH$_4$, N$_2$ and NH$_3$ are determined from the adopted CO/CO$_2$/CH$_3$OH/CH$_4$ and N$_2$/NH$_3$ gas phase molecular ratios, and from the C/O/N relative abundances set in organics. Finally, once the abundances of these molecules are fixed, the remaining O gives the abundance of H$_2$O. We then set CO/CO$_2$/CH$_3$OH/CH$_4$ = 70/10/2/1 in the gas phase of the disk, values that are consistent with the ISM measurements considering the contributions of both gas and solid phases in the lines of sight (Frerking et al. 1982; Ohishi et al. 1992; Ehrenfreund  \& Schutte 2000; Gibb et al. 2000). In addition, S is assumed to exist in the form of H$_2$S, with H$_2$S/H$_2$ = 0.5 $\times$ (S/H$_2$)$_{\odot}$, and other refractory sulfide components 
 (Pasek et al. 2005). We also consider N$_2$/NH$_3$ = 1/1 in the nebula gas-phase. This value is compatible with thermochemical calculations in the solar nebula that take into account catalytic effects of Fe grains on the kinetics of N$_2$ to NH$_3$ conversion (Fegley 2000). In the following, we adopt these mixing ratios as our nominal model of the solar nebula gas phase composition {\bf (see Table \ref{lodders})}.

\begin{figure}
\resizebox{\hsize}{!}{\includegraphics[angle=0]{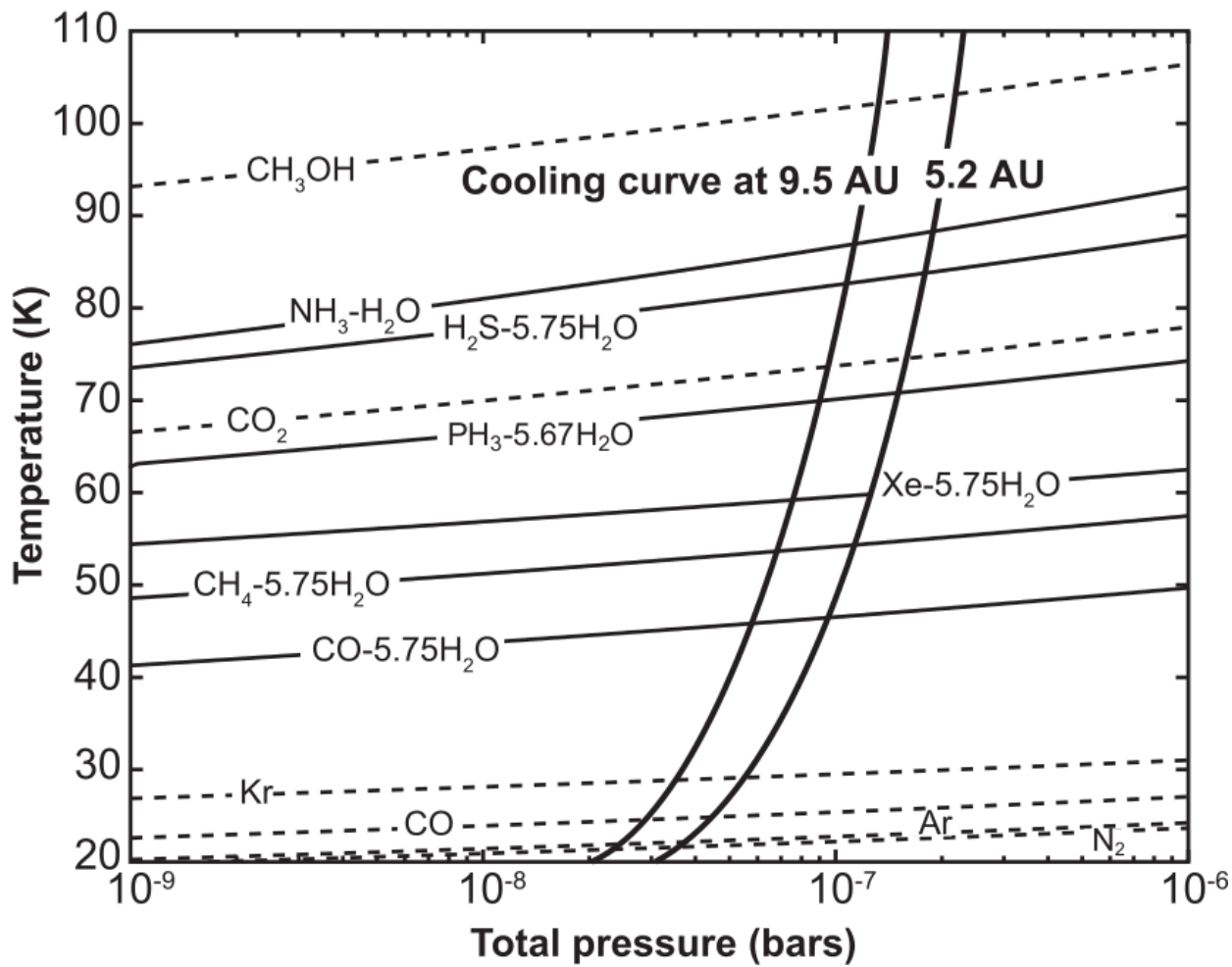}}
\caption{Stability curves of hydrate (NH$_3$-H$_2$O), clathrates (X-5.75H$_2$O or X-5.67H$_2$O) (solid lines), and pure condensates (dotted lines), and cooling curves of the Solar nebula at the heliocentric distances of 5.2 and 9.5 AU, respectively, assuming a full efficiency of clathration. Abundances of various elements are solar, with CO/CO$_2$/CH$_3$OH/CH$_4$ = 70/10/2/1, H$_2$S/H$_2$ = 0.5 $\times$ (S/H$_2$)$_{\odot}$, and N$_2$/NH$_3$ = 1/1 in the gas phase of the disk. Species remain in the gas phase above the stability curves. Below, they are trapped as clathrates or simply condense.}
\label{cool1}
\end{figure}

\begin{figure}
\resizebox{\hsize}{!}{\includegraphics[angle=0]{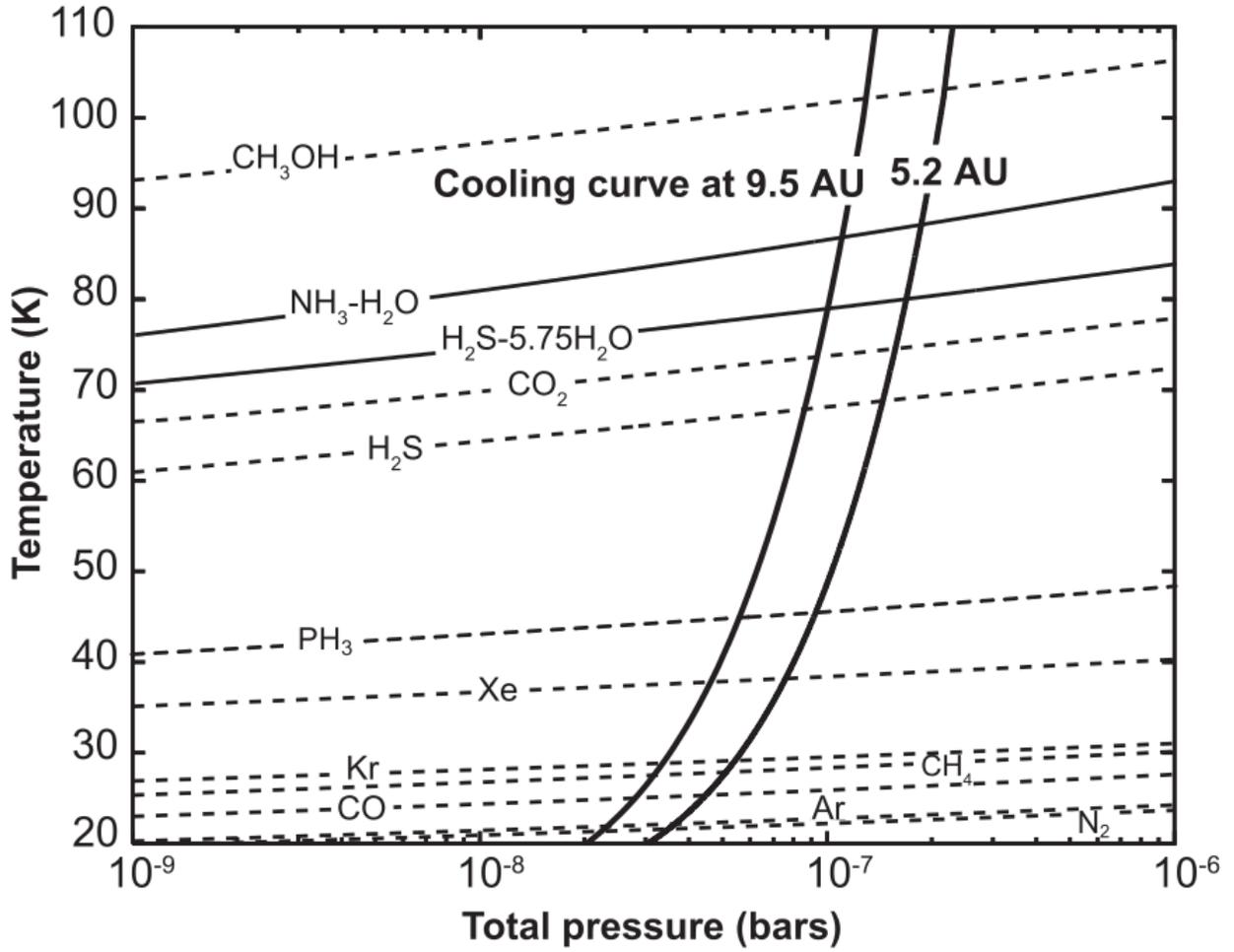}}
\caption{Same as in Fig. 1, but with a 10\% clathration efficiency.}
\label{cool2}
\end{figure}

The process by which volatiles are trapped in icy planetesimals, illustrated in Figs. \ref{cool1} and â\ref{cool2}, is calculated using the stability curves of hydrates, clathrates and pure condensates, and the thermodynamic path detailing the evolution of temperature and pressure at 5.2 and 9.5 AU in the solar nebula, corresponding to the actual positions of Jupiter and Saturn, respectively. We refer the reader to the works of Papaloizou \& Terquem (1999) and Alibert et al. (2005c) for a full description of the turbulent model of accretion disk used here.

The stability curves of hydrates and clathrates derive from Lunine \& Stevenson (1985)'s compilation of published experimental work, in which data are available at relatively low temperatures and pressures. On the other hand, the stability curves of pure condensates used in our calculations derive from the compilation of laboratory data given in the CRC Handbook of Chemistry and Physics (Lide 2002). The cooling curve intercepts the stability curves of the different ices at particular temperatures and pressures. For each ice considered, the domain of stability is the region located below its corresponding stability curve. The clathration process stops when no more crystalline water ice is available to trap the volatile species. Note that, in the pressure conditions of the solar nebula, CO$_2$ is the only species that crystallizes at a higher temperature than its associated clathrate. We then assume that solid CO$_2$ is the only existing condensed form of CO$_2$ in this environment. In addition, we have considered only the formation of pure ice of CH$_3$OH in our calculations since, to our best knowledge, no experimental data concerning the stability curve of its associated clathrate have been reported in the literature. 

Because the clathration efficiency remains unknown in the solar nebula, we have considered a set of sequences of ices formation in which the fraction of water available for clathration is varied between 0 and 100\%. Figure \ref{cool1} illustrates the case where the efficiency of clathration is total, implying that guest molecules had the time to diffuse through porous water-ice planetesimals before their accretion by proto-Jupiter and proto-Saturn. In this case, NH$_3$, H$_2$S, {\bf PH$_3$, Xe, CH$_4$ and $\sim$ 61\% of CO form NH$_3$-H$_2$O hydrate and H$_2$S-5.75H$_2$O, PH$_3$-5.67H$_2$O}, Xe-5.75H$_2$O, CH$_4$-5.75H$_2$O and CO-5.75H$_2$O clathrates with the available water. The remaining CO, as well as N$_2$, Kr, and Ar, whose clathration normally occurs at lower temperatures, remain in the gas phase until the nebula cools enough to allow the formation of pure condensates. Figure \ref{cool2} illustrates the case where the efficiency of clathration is only $\sim$ 10\%. Here, either only a part of the clathrates cages have been filled by guest molecules, either the diffusion of clathrated layers through the planetesimals was to slow to enclathrate most of the ice, or the poor trapping efficiency was the combination of these two processes. In this case, only NH$_3$ and {\bf $\sim$ 21\% of H$_2$S} form NH$_3$-H$_2$O hydrate and H$_2$S-5.75H$_2$O clathrate. Due to the deficiency in accessible water in icy planetesimals, the remaining H$_2$S and all {\bf PH$_3$}, Xe, Kr, CH$_4$, CO, Ar and N$_2$ form pure condensates in the solar nebula.

{\bf Table \ref{trap} summarizes the trapping/formation conditions of the different ices calculated at 5.2 and 9.5 AU in the solar nebula in the cases of 100\% and 10\% clathration efficiencies, and for our nominal gas phase. Using these thermodynamic conditions, one can estimate the mass ratios of the different ices with respect to H$_2$O in the planetesimals formed in the solar nebula. Indeed, the volatile, $i$, to water mass ratio in these planetesimals is determined by the relation given by Mousis \& Gautier (2004):         

\begin{equation}
{m_i = \frac{X_i}{X_{H_2O}} \frac{\Sigma(r; T_i, P_i)}{\Sigma(r; T_{H_2O}, P_{H_2O})}},
\end{equation}

\noindent where $X_i$ and $X_{H_2O}$ are the mass mixing ratios of the volatile $i$ and H$_2$O with respect to H$_2$ in the nebula, respectively. $\Sigma(R; T_i, P_i)$ and $\Sigma(R; T_{H_2O}, P_{H_2O})$ are the surface density of the nebula at a distance $r$ from the Sun at the epoch of hydratation or clathration of the species $i$, and at the epoch of condensation of water, respectively. From ${\it m_i}$, it is possible to determine the mass fraction $M_i$ of species $i$ with respect to all the other volatile species taking part to the formation of an icy planetesimal via the following relation:

\begin{equation}
{M_i = \frac{m_i}{\displaystyle \sum_{j=1,n} m_j}},
\end{equation}

\noindent with $\displaystyle \sum_{i=1,n} M_i = 1$.}

Note that, whatever their formation distance in the nebula, the composition of planetesimals remains almost constant, provided that the gas phase composition does not vary (Marboeuf et al. 2008) {\bf and that the clathration efficiency remains constant. In particular, if they formed in the same gas phase conditions and with the same clathration efficiency in the nebula, the ices accreted by proto-Jupiter and proto-Saturn share a similar composition along their migration pathways. On the other hand, the composition of ices can be somewhat altered by the variation of the clathration efficiency (see Table \ref{comp}). This does not impair the quality of the fits of the observed enrichments but the budget of heavy elements needed in the envelopes of the two planets can be modified (see Sec. \ref{mass}). However, in any case, it appears difficult to provide some constraint on the migration status of Jupiter and Saturn from their observed enrichments because the changes of planetesimals composition are never significant in the giant planets formation zone, even when the clathration efficiency is varied.}

\section{Minimum masses of heavy elements in the envelopes of Jupiter and Saturn}
\label{mass}

From the adjustment of the masses of ices vaporized in the envelopes, we have been able to reproduce the observed volatiles enrichments. {\bf Our strategy was i) to match the maximum number of observed volatiles enrichments and ii) in the range of possible solutions, to select the fit that minimizes the mass of ices needed in the envelopes of Jupiter and Saturn.

In our calculations, we have considered the measurements of C, N, S, Ar, Kr and Xe abundances in Jupiter's atmosphere determined using the mass spectrometer on board the Galileo probe (Mahaffy et al. 2000; Wong et al. 2004).  Estimates of the abundance of P from 10 $\mu$m Cassini/CIRS PH$_3$ observations have been recently updated by Fletcher et al. (2009) to include improved estimates of the mid-IR aerosol opacity, spectroscopic linedata and NH$_3$ abundance, resulting in derived abundances slightly larger than the previous analysis of Irwin et al. (2004). 

In the case of Saturn, we have only used the recent determination of C abundance by Fletcher et al. (2008) from  Cassini CIRS spectra.  This new measurement updates the previous study by Flasar et al. (2005) using thousands of high-resolution spectra in both the mid- and far-infrared acquired during Cassini's prime mission, taken at a range of spatially-resolved locations on the planet.  As well as improving the precision of the measurement, Fletcher et al. (2008) demonstrate the consistency between results obtained from rotational and vibrational line manifolds in the far- and mid-IR, and show the lack of hemispherical asymmetry in the CH$_4$ abundance, confirming the hypothesis that this gas is well-mixed throughout the observable atmosphere and is therefore representative of the bulk composition. Finally, we use a new estimate of Saturn's PH$_3$ abundance, taking into account the latitudinal variability and mid-IR aerosol opacities derived by Fletcher et al. (2009).

We have chosen to omit the mixing ratios of condensible species (NH$_3$, H$_2$S), as infrared remote sensing is unable to constrain the abundances beneath the condensation clouds, and the competing spectral effects of these species are difficult to disentangle from measurements of Saturn's radio frequency opacity (Briggs \& Sackett 1989).   Furthermore, the absence of an atmospheric entry-probe at Saturn means we have no constraints on the abundances of the noble gases (Ar, Kr, Xe) in that planet's atmosphere.}

Table \ref{enri} gives our ``minimum'' fits for the two planets in the case of the nominal model and with 100\% and 10\% clathration efficiencies. {\bf In the case of Saturn and whatever the clathration efficiency, the observed C and P enrichments are systematically matched within error bars or range of values. Similarly, in the case of Jupiter, the observed C, N, S and Ar enrichments are systematically matched within error bars. However, the calculated P abundance in the Jovian atmosphere is higher than the observed value. On the other hand, the relationship between the PH$_3$ abundance observed in the 1--4 bar region on both planets and the P/H content of the interior is dependent upon the assumed value of the eddy mixing coefficient at the kbar level. It is therefore possible that the observed PH$_3$ abundances provide only lower bounds on the P/H abundance (Fegley \& Lodders 1994; Fletcher et al. 2009). Moreover,} the calculated Kr and Xe enrichments are higher than the observed ones but the presence of H$_3^+$ ion in the primitive nebula, which induces an efficient trapping of these species in the form of stable complexes XH$_3^+$ (with X = Kr and Xe; Pauzat \& Ellinger 2007) in the gas phase, may limit their ability to be enclathrated or to condense in the outer nebula (Mousis et al. 2008). Therefore, if the H$_3^+$ abundance were comparable to those of Kr and Xe in Jupiter's feeding zone (H$_3^+$/H$_2$ $\sim$ 10$^{-9}$--10$^{-10}$), which is a reasonable assumption (see e.g. Mousis et al. 2008), the resulting enrichments of these two nobles gases in the Jovian envelope should be lower than the values calculated here and might match the observed values\footnote{Ar also forms stable ArH$_3^+$ complexe in the gas phase (Pauzat \& Ellinger 2007) but its abundance in the solar nebula is several orders-of-magnitude higher than the one envisaged for H$_3^+$. Hence, the fraction of Ar sequestrated by H$_3^+$ can be neglected.}. For example, if H$_3^+$/H$_2$ $\sim$ 5 $\times$ 10$^{-10}$ in the feeding zone of Jupiter and if similar amounts of Kr and Xe have been trapped by H$_3^+$, their corresponding enrichments are now  2.4 and 1.7 and match the observed values in the case of full clathration efficiency. Note that, since the H$_3^+$ abundance is expected to increase with the growing heliocentric distance, the observed deficiency of Titan's atmosphere in Kr and Xe (Niemann et al. 2005) was suggested to be caused by the presence of a higher concentration of KrH$_3^+$ and XeH$_3^+$ complexes in Saturn's feeding zone, inducing the formation of Kr- and Xe-poor planetesimals ultimately accreted by the satellite (Mousis et al. 2008). If this scenario is correct, the abundances of Kr and Xe should be solar in Saturn's atmosphere. In a less extreme case, if we also adopt H$_3^+$/H$_2$ $\sim$ 5 $\times$ 10$^{-10}$ in Saturn's feeding zone, Kr and Xe enrichments should be of 6.8 and 4.7, instead of respectively 7.4 and 11.6 in the case of full clathration efficiency.

\begin{figure}
\resizebox{\hsize}{!}{\includegraphics[angle=0]{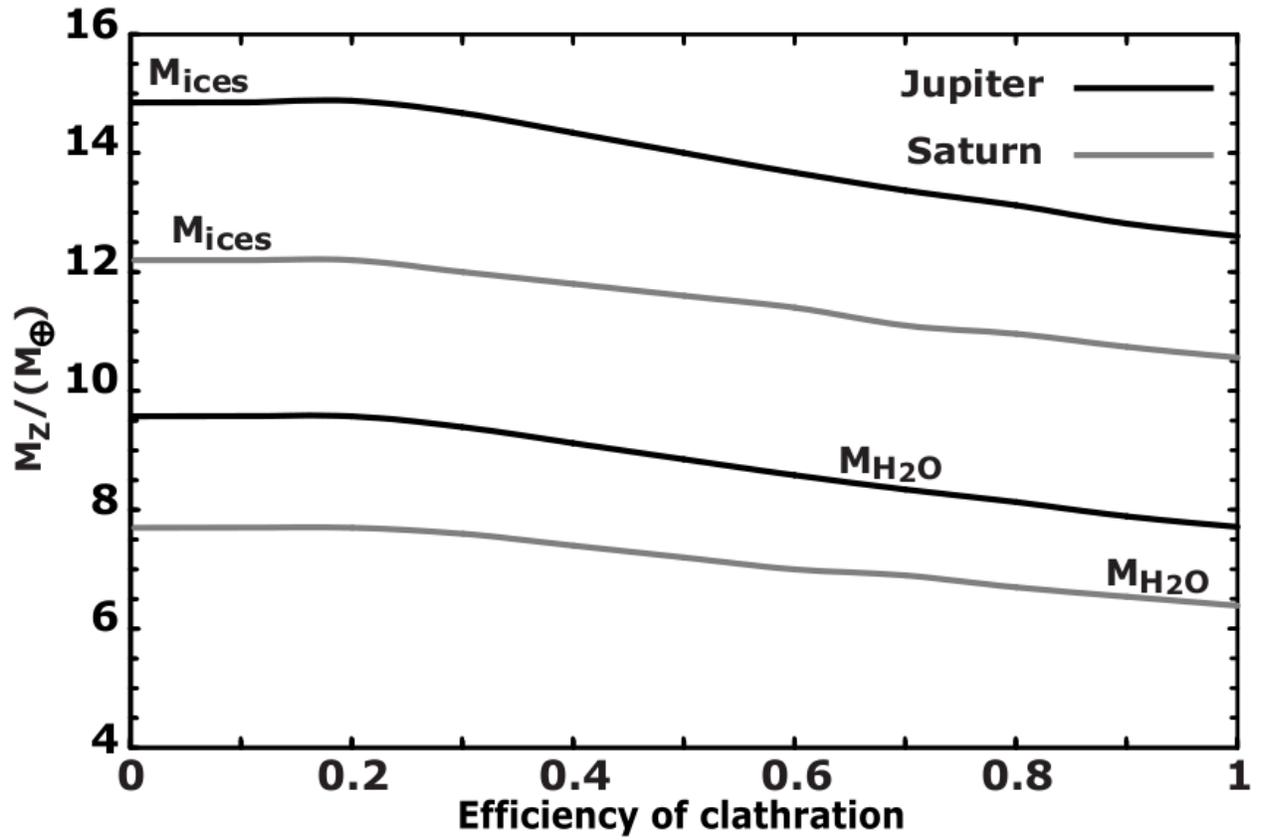}}
\caption{Minimum masses of ices and water needed in the envelopes of Jupiter and Saturn to fit the measured abundances of volatiles in the case of our nominal model, as a function of the clathration efficiency in the two giant planets feeding zones.}
\label{ices1}
\end{figure}

Independently of the efficiency of Xe and Kr trapping by H$_3^+$ in the nebula\footnote{Kr and Xe poorly influence the budget of ices accreted by Jupiter and Saturn, due to their low abundances compared to those of other volatiles.}, our calculations predict that the O/H enrichment decreases from $\sim$ 6.7 to 5.6 times (O/H)$_{\odot}$ in the envelope of Jupiter and from $\sim$ 18.1 to 15.4 times (O/H)$_{\odot}$ in the envelope of Saturn, with the growing clathration efficiency in the solar nebula. Figure \ref{ices1} shows that the minimum masses of ices required in the envelopes of Jupiter and Saturn to match the observed enrichments diminish when the clathration efficiency increases in the solar nebula. This effect illustrates that the variation of the clathration efficiency affects the composition of ices produced in the outer nebula. The minimum mass of ices thus needed to be injected in the envelope of Jupiter decreases from $\sim$ 14.9 to 12.6 $\mearth$, including a mass of water that diminishes from 9.6 to 7.7  $\mearth$, with the growing clathration efficiency. In the same conditions, the minimum mass of ices needed in the envelope of Saturn decreases from $\sim$ 12.2 to 10.6 $\mearth$, including a mass of water that diminishes from 7.7 to 6.4 $\mearth$.

\section{Summary and discussion}
\label{discus}

\begin{itemize}

\item In this report, we have assumed that the water abundance is homogeneous whatever the considered heliocentric distance and derives directly from a gas phase of solar composition. The amount of water available for clathration is then no more a free parameter adjusted in order to ease the fits of the observed volatiles enrichments in Jupiter and Saturn;

\item Planetesimals then form in the feeding zones of Jupiter and Saturn from clathrates and pure condensates in proportions fixed by the clathration efficiency;

\item A fraction of Kr and Xe may have been sequestrated by H$_3^+$ in the form of XeH$_3^+$ and KrH$_3^+$ complexes in the solar nebula gas phase, thus implying the formation of at least partly Xe- and Kr-depleted planetesimals in the feeding zones of Jupiter and Saturn;

\item From plausible molecular mixing ratios in the gas phase of the outer solar nebula and from the calculation of the composition range of ices accreted by the growing Jupiter and Saturn, we show that it is possible to match the observed {\bf C, N, S, Ar, Kr and Xe} enrichments, whatever the clathration efficiency.

\end{itemize}

The minimum mass of ices [12.6--14.9 $\mearth$] required in the envelope of Jupiter to match the observed enrichments is lower than the maximum amount of heavy elements (42 $\mearth$) predicted in the same zone by the internal structure models of SG04. Even if planetesimals accreted by proto-Jupiter are half composed of ices and half made of rocks (ices to rocks ratio of 1), a value compatible with the internal structures of Ganymede and Callisto (Sohl et al. 2002), the global mass of solids injected in the envelope remains lower than the maximum one predicted by SG04. These results are then compatible with formation scenarios of the two Galilean satellites from the accretion of planetesimals formed in the nebula without having been vaporized inside the subdisk (Mousis \& Gautier 2004; Mousis \& Alibert 2006).

On the other hand, the minimum mass of ices [10.6--12.2 $\mearth$] needed in the envelope of Saturn to match the observed enrichments slightly exceeds the maximum amount of heavy elements ($\sim$ 10 $\mearth$) predicted by SG04. However, note that the minimum mass of ices predicted by our model is lower than the [13.7--18.3$\mearth$] range of values derived by Mousis et al. (2006) from the assumption that crystalline water was abundant enough in the feeding zones of Jupiter and Saturn to enclathrate essentially all the volatile species present in the gas phase.

\begin{figure}
\resizebox{\hsize}{!}{\includegraphics[angle=0]{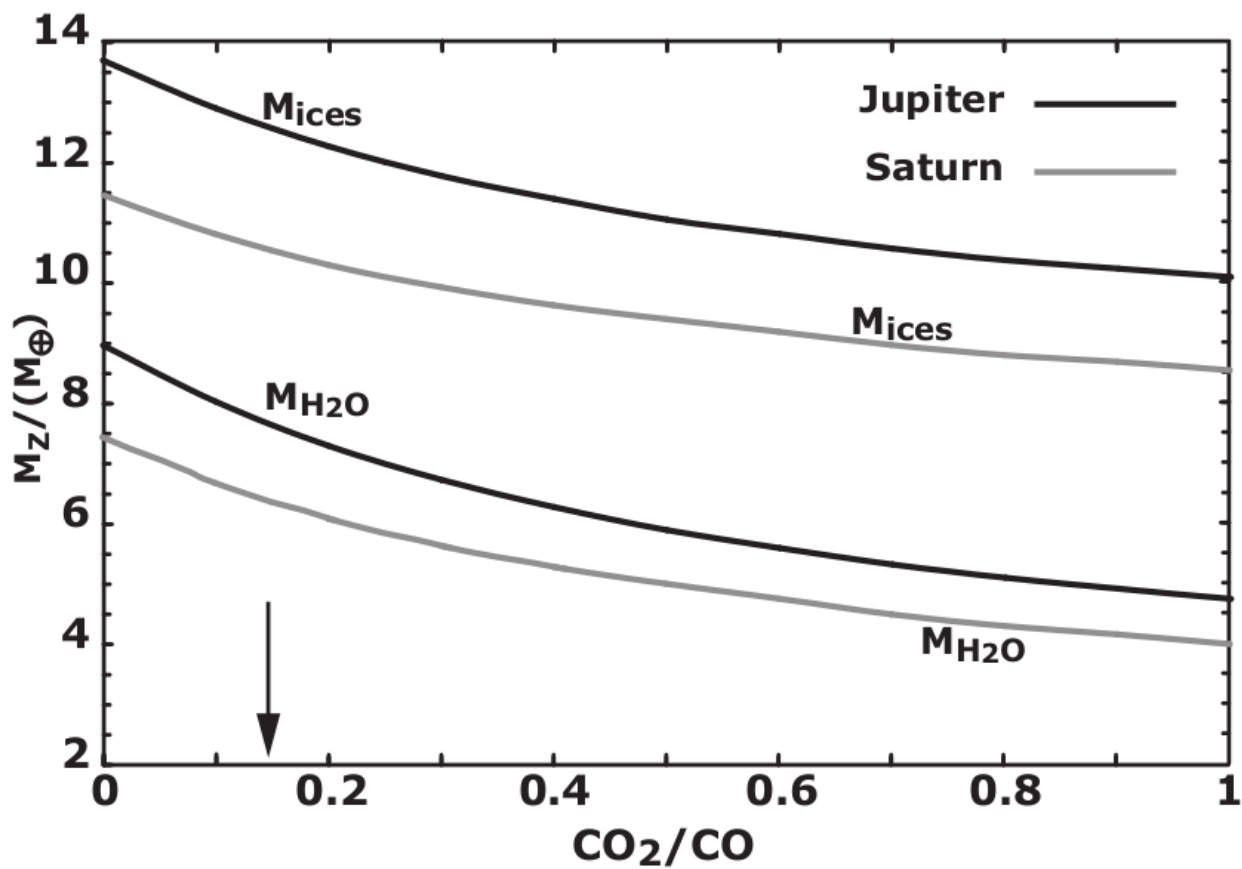}}
\caption{Minimum masses of ices and water needed in the envelopes of Jupiter and Saturn to fit the measured abundances of volatiles, as a function of the CO$_2$/CO ratio postulated in the solar nebula gas phase. Abundances of various elements are solar, with CO/CH$_3$OH/CH$_4$ = 70/2/1, H$_2$S/H$_2$ = 0.5 $\times$ (S/H$_2$)$_{\odot}$, and N$_2$/NH$_3$ = 1/1 in the gas phase of the disk. The vertical arrow indicates the CO$_2$/CO ratio corresponding to our nominal model. Calculations were made with the assumption of a full clathration efficiency.}
\label{ices2}
\end{figure}

It is still possible to slightly decrease the minimum mass of ices needed in the envelopes of Jupiter and Saturn by adopting CO$_2$/CO gas phase ratios different from the one selected in our nominal model. The variation of this ratio rules the distribution of C and O among CO$_2$, CO, CH$_3$OH, CH$_4$ and H$_2$O molecules and then strongly affects the budget of crystalline water available for clathration. Figure \ref{ices2} illustrates this effect and shows that the minimum mass of ices needed in the envelopes of Jupiter and Saturn to match the observed enrichments diminishes when the CO$_2$/CO ratio increases in the initial gas phase of disk. In the case of Saturn, CO$_2$/CO ratios greater than $\sim$ 0.2 in the nebula gas phase allow the minimum mass of ices needed in the envelope to be lower than the maximum value given by SG04. However, it is difficult to explain the accretion in Saturn of planetesimals with ice to rock ratios $\sim$ 1 --similar to that predicted in Titan by internal structure models (Tobie et al. 2006)-- by invoking higher CO$_2$/CO ratios in the nebulae gas phase, in agreement with the amount of ices predicted in the envelope by SG04. 

Indeed, the extremes adopted in Fig. \ref{ices2} for the CO$_2$/CO ratio in the nebula --0.1-1.0-- correspond to contributions from, respectively, the ISM gas and solid phases (Ehrenfreund \& Schutte 2000; Gibb et al. 2004).  Hence, any CO$_2$/CO mixing ratio adopted in the solar nebula gas phase should hold within this range of values. On the other hand, if a substantial fraction of rock contained in planetesimals that accreted in the envelope has sedimented onto the core of Saturn during its evolution\footnote{Due to their higher density and their low volatility, rock should reach the deepest layers of Saturn's envelope during the accretion of planetesimals and remain in the solid phase at extremely high pressure.}, the inconsistency between the accreted mass of planetesimals and the one predicted by interior models could be removed.

{\bf Here, we have assumed that the observed enrichments in volatiles were engendered by the vaporization of icy planetesimals when they entered the envelopes of the growing planets. On the other hand, the erosion of a significant part of the giant planets cores could also constitute a possible source of volatiles enrichments in their atmospheres. However, in this case, the solids accreted by the cores should share the same composition as those accreted later in the envelope. Hence, the approach described in this work remains valid whatever the delivery process of volatiles (core erosion or accretion of solids in the envelope).}

Finally, we note that a new compilation of protosolar abundances has been recently published by Grevesse et al. (2007). Comparisons between this compilation and the one used in the present work and taken from Lodders (2003) show that, except for Ar abundance which is $\sim$ 2.5 times smaller in the more recent tabulation, {\bf no such substantial variation is observed. If Ar is not considered, the use of the compilation of Grevesse et al. (2007) do not alter the conclusions of this paper.} However, if we consider the Ar abundance given by Grevesse et al. (2007), only the observed C, N and S enrichments in Jupiter can be matched by our calculations, whatever the clathration efficiency. The use of two different methods seems to be at the origin of the difference in the quoted solar Ar abundance in these compilations. Indeed, Lodders (2003) based the Ar abundance on nucleosynthesis arguments whereas Grevesse et al. (2007) based the Ar abundance from abundance ratios of Ar to other elements measured in the solar wind (SW) and solar energetic particles (SEPs) coming from the Sun's corona. Following Lodders (2008), the limitation of the method employed by Grevesse et al. (2007) is that elements may become fractionated in the SW and SEPs from photospheric abundances according to their different first ionization potentials (FIP). Relative to photospheric values, elements with low FIP $<$ 10 eV would be enriched in the SW and SEPs, whereas elements with high FIP (such as Ar) appear to be depleted (Lodders 2003,2008). From these considerations, and for reasons of consistency, we have adopted the compilation of Lodders (2003) in all our calculations.

\acknowledgments
O.M. and U. M. acknowledge the financial support of the ANR HOLMES. This work was supported in part by the French Centre National d'Etudes Spatiales. JL was supported by the Juno Project at the Southwest Research Institute. LF was supported by an appointment to the NASA Postdoctoral Program at the Jet Propulsion Laboratory, administered by Oak Ridge Associated Universities through a contract with NASA. Tristan Guillot is acknowledged for helpful discussions.

\clearpage

\begin{table}
\caption[]{Elemental and molecular abundances in the solar nebula}
\begin{center}
\begin{tabular}{lc}
\hline
\hline
\noalign{\smallskip}
Species X 	&  X/H$_2$ \\	
\noalign{\smallskip}
\hline
\noalign{\smallskip}
C			& $5.82 \times 10^{-4}$ 		\\
N   			& $1.60 \times 10^{-4}$    		\\
O			& $1.16 \times 10^{-3}$		\\
S        		& $3.66 \times 10^{-5}$		\\
Ar       		& $8.43 \times 10^{-6}$		\\
Kr       		& $4.54 \times 10^{-9}$		\\
Xe       		& $4.44 \times 10^{-10}$		\\
P			& $6.88 \times 10^{-7}$		\\
H$_2$O  		& $4.44 \times 10^{-4}$		\\
CO      		& $2.21 \times 10^{-4}$		\\
CO$_2$  		& $3.16 \times 10^{-5}$		\\
NH$_3$   		& $4.05 \times 10^{-5}$		\\
H$_2$S    	& $1.83 \times 10^{-5}$		\\
N$_2$		& $4.05 \times 10^{-5}$		\\
CH$_3$OH  	& $6.31 \times 10^{-6}$		\\ 
CH$_4$  		& $3.16 \times 10^{-6}$		\\
PH$_3$		& $6.88 \times 10^{-7}$		\\   			  
Kr       		& $4.54 \times 10^{-9}$		\\	
Xe       		& $4.44 \times 10^{-10}$		\\
\hline
\end{tabular}
\end{center}
\tablecomments{Elemental abundances derive from Lodders (2003). Molecular abundances are determined from our nominal gas phase composition.}
\label{lodders}
\end{table}

\clearpage

\begin{table*}
\caption[]{Thermodynamic conditions of ices formation in the outer nebula}
\begin{center}
\begin{tabular}{lcccccc}
\hline
\hline
\noalign{\smallskip}
\multicolumn{7}{c}{100\% clathration efficiency}		 \\
\noalign{\smallskip}
\hline
\noalign{\smallskip}
Heliocentric distance (AU)	&	\multicolumn{3}{c}{5.2} 	&	\multicolumn{3}{c}{9.5} \\
\noalign{\smallskip}
\hline
\noalign{\smallskip}
Ice 		 				& $T$ (K)	& $P$ (bars)	& $\Sigma$ (g.cm$^{-2}$)	& $T$ (K)	& $P$ (bars)	& $\Sigma$ (g.cm$^{-2}$)\\	
\noalign{\smallskip}
\hline
\noalign{\smallskip}
H$_2$O					& 155.65 	& $3.35  \times 10^{-7}$	& 670.73	& 155.51 	& $2.03  \times 10^{-7}$	& 1146.91	\\
CH$_3$OH				& 103.18	& $2.18  \times 10^{-7}$	& 536.40	& 104.35 	& $1.32  \times 10^{-7}$	& 915.31	\\
NH$_3$-H$_2$O     			& 88.31	& $1.87  \times 10^{-7}$	& 495.52	& 88.65 	& $1.13  \times 10^{-7}$	& 845.97	\\
H$_2$S-5.75H$_2$O		& 83.89	& $1.77  \times 10^{-7}$	& 482.24	& 85.23 	& $1.08  \times 10^{-7}$	& 829.15	\\
CO$_2$       				& 74.55 	& $1.57  \times 10^{-7}$	& 452.59	& 75.76 	& $9.62  \times 10^{-8}$	& 780.15	\\
PH$_3$-5.67H$_2$O		& 70.84	& $1.49  \times 10^{-7}$	& 440.10	& 70.05 	& $8.89  \times 10^{-8}$	& 748.65	\\
Xe-5.75H$_2$O			& 59.83	& $1.25  \times 10^{-7}$	& 400.05	& 59.95 	& $7.58  \times 10^{-8}$	& 688.23 	\\
CH$_4$-5.75H$_2$O   		& 54.37     & $1.13  \times 10^{-7}$	& 378.12	& 55.49 	& $6.99  \times 10^{-8}$	& 659.24 	\\
CO-5.75H$_2$O        		& 47.46	& $9.72  \times 10^{-8}$	& 347.84	& 47.59 	& $5.95  \times 10^{-8}$	& 603.46	\\
Kr						& 29.15	& $5.43  \times 10^{-8}$	& 244.10	& 29.46 	& $3.45  \times 10^{-8}$	& 439.45	\\
CO        					& 25.33	& $4.47  \times 10^{-8}$	& 214.44	& 25.67 	& $2.90  \times 10^{-8}$	& 393.47	\\
Ar						& 22.17	& $3.64  \times 10^{-8}$	& 185.64	& 22.53  	& $2.42  \times 10^{-8}$	& 349.41 	\\
N$_2$       				& 21.60	& $3.49  \times 10^{-8}$	& 179.95	& 21.83 	& $2.32  \times 10^{-8}$	& 338.70	\\
\noalign{\smallskip}
\hline
\noalign{\smallskip}
\multicolumn{7}{c}{10\% clathration efficiency}		 \\
\noalign{\smallskip}
\hline
\noalign{\smallskip}
Heliocentric distance (AU)	&	\multicolumn{3}{c}{5.2} 	&	\multicolumn{3}{c}{9.5} \\
\noalign{\smallskip}
\hline
\noalign{\smallskip}
Ice		 				& $T$ (K)	& $P$ (bars)	& $\Sigma$ (g.cm$^{-2}$)	& $T$ (K)	& $P$ (bars)	& $\Sigma$ (g.cm$^{-2}$)\\	
\noalign{\smallskip}
\hline
\noalign{\smallskip}
H$_2$O					& 155.65 	& $3.35  \times 10^{-7}$	& 670.73	& 155.51 	& $2.03  \times 10^{-7}$	& 1146.91	\\
CH$_3$OH				& 103.18	& $2.18  \times 10^{-7}$	& 536.40	& 104.35 	& $1.32  \times 10^{-7}$	& 915.31	\\
NH$_3$-H$_2$O     			& 88.31	& $1.87  \times 10^{-7}$	& 495.52	& 88.65 	& $1.13  \times 10^{-7}$	& 845.97	\\
H$_2$S-5.75H$_2$O		& 83.89	& $1.77  \times 10^{-7}$	& 482.24	& 85.23 	& $1.08  \times 10^{-7}$	& 829.15	\\
CO$_2$       				& 74.55 	& $1.57  \times 10^{-7}$	& 452.59	& 75.76 	& $9.62  \times 10^{-8}$	& 780.15	\\
H$_2$S       				& 68.80 	& $1.45  \times 10^{-7}$	& 433.01	& 70.05 	& $8.89  \times 10^{-8}$	& 748.65	\\
PH$_3$					& 45.52	& $9.28  \times 10^{-8}$	& 338.73	& 45.81 	& $5.71  \times 10^{-8}$	& 589.95	\\
Xe						& 38.20	& $7.59  \times 10^{-8}$	& 301.06	& 37.96 	& $4.64  \times 10^{-8}$	& 524.77 	\\
Kr						& 29.15	& $5.43  \times 10^{-8}$	& 244.10	& 29.46 	& $3.45  \times 10^{-8}$	& 439.45	\\
CH$_4$			   		& 27.87     & $5.11  \times 10^{-8}$	& 234.69	& 28.44 	& $3.30  \times 10^{-8}$	& 427.78 	\\
CO        					& 25.33	& $4.47  \times 10^{-8}$	& 214.44	& 25.67 	& $2.90  \times 10^{-8}$	& 393.47	\\
Ar						& 22.17	& $3.64  \times 10^{-8}$	& 185.64	& 22.53  	& $2.42  \times 10^{-8}$	& 349.41 	\\
N$_2$       				& 21.60	& $3.49  \times 10^{-8}$	& 179.95	& 21.83 	& $2.32  \times 10^{-8}$	& 338.70	\\
\hline
\end{tabular}
\end{center}
\tablecomments{$T$, $P$ and $\Sigma$ are the temperature, pressure and surface density of the H$_2$-dominated gas given at 5.2 or 9.5 AU in the solar nebula at the epoch of condensation of volatile $i$ or its trapping by crystalline water ice.}
\label{trap}
\end{table*}

\clearpage

\begin{table}
\caption[]{Average composition of ices formed in the outer solar nebula}
\begin{center}
\begin{tabular}{lcc}
\hline
\hline
\noalign{\smallskip}
Ice		 	&  100\% clathration efficiency & 10\% clathration efficiency\\	
\noalign{\smallskip}
\hline
\noalign{\smallskip}
H$_2$O			& $6.08 \times 10^{-1}$ 		& $6.39 \times 10^{-1}$	\\
CO        			& $2.00 \times 10^{-1}$		& $1.64 \times 10^{-1}$	\\
CO$_2$       		& $7.18 \times 10^{-2}$		& $7.54 \times 10^{-2}$	\\
NH$_3$       		& $3.88 \times 10^{-2}$		& $4.07 \times 10^{-2}$	\\
H$_2$S			& $3.42 \times 10^{-2}$		& $3.30 \times 10^{-2}$	\\
N$_2$       		& $2.44 \times 10^{-2}$		& $2.56 \times 10^{-2}$	\\
CH$_3$OH		& $1.23 \times 10^{-2}$		& $1.29 \times 10^{-2}$	\\
Ar				& $6.73 \times 10^{-3}$		& $7.07 \times 10^{-3}$	\\
CH$_4$   			& $2.19 \times 10^{-3}$    		& $1.46 \times 10^{-3}$	\\
PH$_3$			& $1.17 \times 10^{-3}$		& $9.55 \times 10^{-4}$	\\
Kr				& $1.08 \times 10^{-5}$		& $1.14 \times 10^{-5}$	\\
Xe				& $2.66 \times 10^{-6}$		& $2.11 \times 10^{-6}$	\\
\hline
\end{tabular}
\end{center}
\tablecomments{Composition of ices is calculated in the cases of 100\% and 10\% clathration efficiencies. Ratio of the mass of ice ${\it i}$ to the global mass of ices (wt\%) is determined from our nominal gas phase composition.}
\label{comp}
\end{table}

\clearpage

\begin{table}
\caption[]{Observed enrichments in volatiles in Jupiter and Saturn, and calculated enrichments in the case of our nominal model}
\begin{center}
\begin{tabular}{lcccccc}
\hline
\hline
\noalign{\smallskip}
&	\multicolumn{3}{l}{Jupiter}	& \multicolumn{3}{l}{Saturn} \\
Species 	&  Observed 			& (1) 	& (2)   		&  Observed 		& (1) 	& (2) 	\\
\noalign{\smallskip}
\hline
\noalign{\smallskip}
O		&					&  5.6	& 6.7		&					& 15.4		& 18.1	\\
C   		&    	$4.1 \pm 1^a$        	&  3.1	&  3.1	&  	$9.2 \pm 0.4^d$	& 8.8			&  8.8 	\\
N   		&    	$4.15 \pm 1.6^a$ 	&  3.0	&  3.7	&  					& 8.6			&  10.4	\\
S   		&    	$2.4 \pm 0.6^a$    	&  2.5 	&  2.9	&  					& 7.0			&  7.8	\\
P		&	$3.2 \pm 0.15^b$	&  4.6	&  4.4	& 	8.9 -- 13.5$^b$		& 12.6		& 12.1	\\
Ar  		&    	$2.15 \pm 0.4^c$ 	&  1.9	&  2.4	&					& 5.9			&  7.1	\\
Kr   		&    	$2 \pm 0.4^c$       	&  2.6	&  3.2	&					& 7.4			&  9.0	\\
Xe  		&   	$2 \pm 0.4^c$       	&  4.2 	&  3.9	&  					& 11.6		&  10.7	\\
\hline
\end{tabular}
\end{center}
\tablecomments{Cases (1) and (2) correspond to 100\% and 10\% clathration efficiencies, respectively. The sequestration of Kr and Xe in the forms of KrH$_3^+$ and XeH$_3^+$ in the solar nebula gas phase is not taken into account in the presented calculations (see text). The observed values are derived from $^a$ Wong et al. 2004, $^b$ Fletcher et al. (2009), $^c$ Mahaffy et al. (2000) and $^d$ Fletcher et al. (2008), using the protosolar abundances of Lodders (2003).}
\label{enri}
\end{table}

\clearpage

\clearpage

\clearpage

\clearpage

\begin{thebibliography}{}

\bibitem[Alibert et al.(2005)]{2005ApJ...622L.145A} Alibert, Y., Mousis, O., \& Benz, W.\ 2005a, \apjl, 622, L145 

\bibitem[Alibert et al.(2005)]{2005ApJ...626L..57A} Alibert, Y., Mousis, O., Mordasini, C., \& Benz, W.\ 2005b, \apjl, 626, L57

\bibitem[Alibert et al.(2005)]{2005A&A...434..343A} Alibert, Y., Mordasini, C., Benz, W., \& Winisdoerffer, C.\ 2005c, \aap, 434, 343

\bibitem[Briggs \& Sackett(1989)]{1989Icar...80...77B} Briggs, F.~H., \& Sackett, P.~D.\ 1989, Icarus, 80, 77

\bibitem[Ehrenfreund \& Schutte(2000)]{2000AdSpR..25.2177E} Ehrenfreund, P., \& Schutte, W.~A.\ 2000, Advances in Space Research, 25, 2177 

\bibitem[Fegley(2000)]{2000SSRv...92..177F} Fegley, B.~J.\ 2000, Space Science Reviews, 92, 177 

\bibitem[Fegley \& Lodders(1994)]{1994Icar..110..117F} Fegley, B.~J., \& Lodders, K.\ 1994, Icarus, 110, 117

\bibitem[Flasar et al.(2005)]{2005Sci...308..975F} Flasar, F.~M., et al.\ 2005, Science, 308, 975

\bibitem[\protect\citeauthoryear{Fletcher et al.}{2008}]{2008Sci...308..975F} Fletcher L.~N., Orton, G. S., Teanby, N. A., Irwin, P. G. J., \& Bjoraker, G. L., 2008, Icarus, in press

{\bf \bibitem[\protect\citeauthoryear{Fletcher et al.}{2009}]{2009Sci...308..975F} Fletcher, L. N., Orton, G. S., Teanby, N. A., \& Irwin, P. G. J. \ 2009, Icarus, submitted}

\bibitem[Frerking et al.(1982)]{1982ApJ...262..590F} Frerking, M.~A., Langer, W.~D., \& Wilson, R.~W.\ 1982, \apj, 262, 590

\bibitem[Gautier et al.(2001)]{2001ApJ...559L.183G} Gautier, D., Hersant, F., Mousis, O., \& Lunine, J.~I.\ 2001, \apjl, 559, L183 

\bibitem[Gautier et al.(2001)]{2001ApJ...550L.227G} Gautier, D., Hersant, F., Mousis, O., \& Lunine, J.~I.\ 2001, \apjl, 550, L227 

\bibitem[Gibb et al.(2004)]{2004ApJS..151...35G} Gibb, E.~L., Whittet, D.~C.~B., Boogert, A.~C.~A., \& Tielens, A.~G.~G.~M.\ 2004, \apjs, 151, 35

\bibitem[Gibb et al.(2000)]{2000ApJ...536..347G} Gibb, E.~L., et al.\ 2000, \apj, 536, 347

\bibitem[Grevesse et al.(2007)]{2007SSRv..130..105G} Grevesse, N., Asplund, M., \& Sauval, A.~J.\ 2007, Space Science Reviews, 130, 105

\bibitem[Hersant et al.(2008)]{2008P&SS...56.1103H} Hersant, F., Gautier, D., Tobie, G., \& Lunine, J.~I.\ 2008, \planss, 56, 1103 

\bibitem[Hersant et al.(2004)]{2004P&SS...52..623H} Hersant, F., Gautier, D., \& Lunine, J.~I.\ 2004, \planss, 52, 623

\bibitem[Hubickyj et al.(2005)]{2005Icar..179..415H} Hubickyj, O., Bodenheimer, P., \& Lissauer, J.~J.\ 2005, Icarus, 179, 415

{\bf \bibitem[Irwin et al.(2004)]{2004Icar..172...37I} Irwin, P.~G.~J., Parrish, P., Fouchet, T., Calcutt, S.~B., Taylor, F.~W., Simon-Miller, A.~A., \& Nixon, C.~A.\ 2004, Icarus, 172, 37}

\bibitem[Jessberger et al.(1988)]{1988Natur.332..691J} Jessberger, E.~K., Christoforidis, A., \& Kissel, J.\ 1988, \nat, 332, 691 

\bibitem[Lide(2002)]{2002crc..book.....L} Lide, D.~R.\ 2002, CRC Handbook of chemistry and physics : a ready-reference book of chemical and physical data, 83rd ed., by David R.~Lide.~Boca Raton: CRC Press, ISBN 0849304830, 2002

\bibitem[Lodders(2008)]{2008ApJ...674..607L} Lodders, K.\ 2008, \apj, 674, 607 

\bibitem[Lodders(2003)]{2003ApJ...591.1220L} Lodders, K.\ 2003, \apj, 591, 1220

\bibitem[Lunine \& Stevenson(1985)]{1985ApJS...58..493L} Lunine, J.~I., \& Stevenson, D.~J.\ 1985, \apjs, 58, 493 

\bibitem[Mahaffy et al.(2000)]{2000JGR...10515061M} Mahaffy, P.~R., Niemann, H.~B., Alpert, A., Atreya, S.~K., Demick, J., Donahue, T.~M., Harpold, D.~N., \& Owen, T.~C.\ 2000, \jgr, 105, 15061

\bibitem[Marboeuf et al.(2008)]{2008ApJ...681.1624M} Marboeuf, U., Mousis, O., Ehrenreich, D., Alibert, Y., Cassan, A., Wakelam, V., \& Beaulieu, J.-P.\ 2008, \apj, 681, 1624 

\bibitem[Mousis et al.(2008)]{2008ApJ...673..637M} Mousis, O., Pauzat, F., Ellinger, Y., \& Ceccarelli, C.\ 2008, \apj, 673, 637 

\bibitem[Mousis et al.(2006)]{2006A&A...449..411M} Mousis, O., Alibert, Y., \& Benz, W.\ 2006, \aap, 449, 411 

\bibitem[Mousis \& Alibert(2006)]{2006A&A...448..771M} Mousis, O., \& Alibert, Y.\ 2006, \aap, 448, 771

\bibitem[Mousis \& Gautier(2004)]{2004P&SS...52..361M} Mousis, O., \& Gautier, D.\ 2004, \planss, 52, 361 

\bibitem[Mordasini et al.(2008)]{2008A&A...449..411M} Mordasini, C., Alibert, Y., \& Benz, W.\ 2008, \aap, submitted

\bibitem[Niemann et al.(2005)]{2005Natur.438..779N} Niemann, H.~B., et al.\ 2005, \nat, 438, 779

\bibitem[Ohishi et al.(1992)]{1992IAUS..150..171O} Ohishi, M., Irvine, W.~M., \& Kaifu, N.\ 1992, Astrochemistry of Cosmic Phenomena, 150, 171

\bibitem[Owen \& Bar-Nun(2000)]{2000orem.book..459O} Owen, T.~C., \& Bar-Nun, A.\ 2000, Origin of the earth and moon, edited by R.M.~Canup and K.~Righter and 69 collaborating authors.~Tucson: University of Arizona Press., p.459-471, 459 

\bibitem[Owen et al.(1999)]{1999Natur.402..269O} Owen, T., Mahaffy, P., Niemann, H.~B., Atreya, S., Donahue, T., Bar-Nun, A., \& de Pater, I.\ 1999, \nat, 402, 269 

\bibitem[Papaloizou \& Terquem(1999)]{1999ApJ...521..823P} Papaloizou, J.~C.~B., \& Terquem, C.\ 1999, \apj, 521, 823 

\bibitem[Pasek et al.(2005)]{2005Icar..175....1P} Pasek, M.~A., Milsom, J.~A., Ciesla, F.~J., Lauretta, D.~S., Sharp, C.~M., \& Lunine, J.~I.\ 2005, Icarus, 175, 1

\bibitem[Pauzat \& Ellinger(2007)]{2007JChPh.127a4308P} Pauzat, F., \& Ellinger, Y.\ 2007, \jcp, 127, 014308

\bibitem[Pollack et al.(1994)]{1994ApJ...421..615P} Pollack, J.~B., Hollenbach, D., Beckwith, S., Simonelli, D.~P., Roush, T., \& Fong, W.\ 1994, \apj, 421, 615

\bibitem[Saumon \& Guillot(2004)]{2004ApJ...609.1170S} Saumon, D., \& Guillot, T.\ 2004, \apj, 609, 1170

\bibitem[Sohl et al.(2002)]{2002Icar..157..104S} Sohl, F., Spohn, T., Breuer, D., \& Nagel, K.\ 2002, Icarus, 157, 104

\bibitem[Tobie et al.(2006)]{2006Natur.440...61T} Tobie, G., Lunine, J.~I., \& Sotin, C.\ 2006, \nat, 440, 61 

\bibitem[Wong et al.(2004)]{2004Icar..171..153W} Wong, M.~H., Mahaffy, P.~R., Atreya, S.~K., Niemann, H.~B., \& Owen, T.~C.\ 2004, Icarus, 171, 153

\end{thebibliography}
\end{document}